# Some collaboration-competition bipartite networks

Xiu-Lian Xu, Chun-Hua Fu, Dan Shen, Ai-Fen Liu, Da-Ren He

*College of Physics Science and Technology, Yangzhou University, Yangzhou, 225002, China*

**Abstract** Recently, we introduced a quantity, "node weight", to describe the collaboration sharing or competition gain of the elements in the collaboration-competition networks, which can be well described by bipartite graphs. We find that the node weight distributions of all the networks follow the so-called "shifted power law (SPL)". The common distribution function may indicate that the evolution of the collaboration and competition in very different systems obeys a general rule. In order to set up a base of the further investigations on the universal system evolution dynamics, we now present the definition of the networks and their node weights, the node weight distributions, as well as the evolution durations of 15 real world collaboration-competition systems which are belonging to diverse fields.

## 1. Introduction

Studies of complex network appeared as a frontier of statistical physics since the end of last century [1-2]. During recent years, the scientists in many fields revealed a number of common properties for the systems in technological, biological, ecological, economical, social, and many other complex systems by using the concepts and methods of complex networks. In the complex systems, the elements often tend to collaborate and/or compete with each other. Recently, this kind of collaboration-competition networks attracted much attention, and the bipartite graphs are shown to be a useful tool for the description of this kind of collaboration-competition networks. [2-10] A bipartite graph contains two types of nodes, one is called "acts", which can be organizations, events, or activities, and the other is called "actors", which are participates. Only different types of nodes in the bipartite graphs can be connected by edges. To describe the collaboration-competition relation between the actors, a projected single-mode (unipartite) network is often used. In the unipartite network, all the actors, which take part in a same act, are connected by equivalent unweighted links.

We are concentrating on such systems where actors are simultaneously cooperating and competing in acts. In each act some actors make concerted effort to accomplish a task and often create a type of production. The production often induces several kinds of resources. The actors, when they are cooperating, are also competing for a larger piece of the resources. For example, some Hollywood actors work together to produce a movie, which should bring ticket office income and famousness. The former is countable, while the later is not. In most of the previous investigations, where the Hollywood actor collaboration networks were studied, only the collaboration relationship was considered. [1-9] In order to describe both the collaboration and competition, we defined a new quantity "node weight". [10,11]

Basically a node weight was defined as the part of a countable resource, which an actor shares. However, in a more dissectional consideration, the actual competition intensity of two actors should depend on the number of actors taking part in their common act, which is addressed as the "act size" and is defined as $T_l = \sum_i b_{il}$. [7-11] In the definition the bipartite adjacency matrix element $b_{il}$ is defined as $b_{il}$=1 if actor node $i$ and act node $l$ are connected in the bipartite graph; and $b_{il}$=0 in other case. According to this consideration the node weight was defined as follows. If $X_l$ denotes the total countable resource in act $l$, and $x_{il}$ denotes the part shared by actor $i$, the weight of node $i$ in act $l$ was defined as $T_l x_{il}$. The normalized node total weight (NNTW) $\omega_i$ was defined as $\omega_i = (\sum_l T_l x_{il}) / \sum_j [(\sum_l T_l x_{jl})]$.[11]

Under this definition it is interesting to know the distribution function of the NNTW, $\omega_i$, which should let us know the disparity of the competition gain. In the rest sections we will present the definition of 14 real world collaboration-competition networks and their node weights, as well as the empirical NNTW distributions. We find that all the NNTW distributions follow the so-called "shifted power law (SPL)", which can be expressed as $P(x) \propto (x+\alpha)^{-\gamma}$.[7-11] When $\alpha = 0$, it takes a power law form. In the condition that the $x$ is normalized ($0 < x_i < 1$ and $\sum_{i=1}^{N} x_i = 1$), we can prove (see appendix 1) that SPL function tends to an exponential function if $\alpha \to 1$. Therefore an SPL interpolates between a power law and an exponential function and the parameter $\alpha$ characterizes the degree of departure from a power law. The parameter $\gamma$ signifies the scaling exponent when SPL approaches to a power form.[7,11] The common distribution function may indicate that the evolution of the collaboration and competition in very different systems obeys a general rule.

In order to set up a base of the further investigations on the universal system evolution dynamics, we also present the empirical evolution durations of the 10 real world systems, since the observed node weight distributions are results of actor's collaboration-competition in the duration. In most cases, the collaboration/competition between the actors occurs right after the birth of the acts, therefore, we can define the time duration as the so-called "act duration" (namely, the time duration between the act birth and the act termination or data collection). Here we define the "act birth" as the time when the first actor joins it, and the "act termination" as the time when the collaboration task is accomplished and the actors disband.

The description of the dynamics, which we shall report in a future paper, certainly will be in a logarithmic time scale because the act durations of different systems must be very different and cross many order of magnitudes. We emphasize that, for the purpose, it is enough to get the reliable data of the act duration order of magnitudes. Although for some systems the exact duration data of all the acts can be obtained, such a high accuracy is useless for the purpose. Also, usually there are many acts in each system and the acts show very different act durations. It is meaningless to list all the act duration data; instead, we shall only present the order of magnitudes of the longest and the shortest ones.

## 2. 2004 Athens Olympic network

In an Olympic game some athletes join a sport event to successfully conduct the pageant and also to obtain more sport scores. We construct the network of the 2004 Athens Olympic Game by defining the athletes as the actor nodes and the sport events (only the individual sport events, e.g., high jump, weight lifting, are considered) as acts. The data were downloaded from www.sina.com.cn (2004), which includes 133 individual sport events and 4500 athletes, as well as their sport scores in each sport event.[12-15] An edge in the bipartite graph represents that an athlete takes part in an event. In the projected unipartite graph two actors are connected if they join in at least one common act. In order to reduce the unavoidable statistical fluctuation, Fig. 1a shows the empirical cumulative distribution of the NNTW, $P(\omega'\geq\omega)$, for the 2004 Athens Olympic network. The fitting solid line denotes the most suitable shifted power law function. The fitted parameters are $\alpha=1$ and $\gamma=3617$. Obviously, the SPL function regresses to an exponential function because $\alpha=1$. The corresponding fitting of exponential function is shown in Fig.1b.

All the act duration data can be downloaded. Within the acts, hectometer match shows the shortest act duration, which is 10.1 seconds. In comparison, Triathlon shows the longest act duration which is 6 hours and 25 seconds. The average of the act durations for this network is around 2 hours.

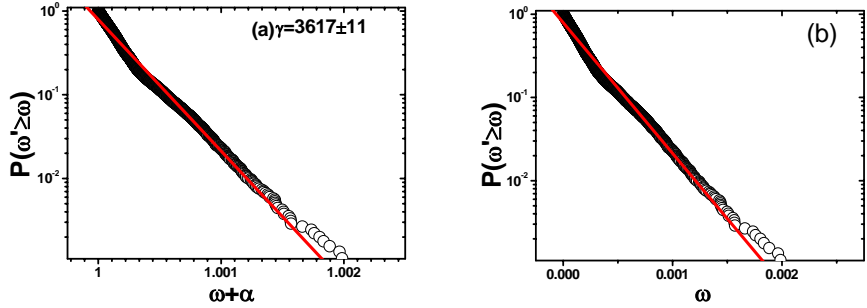

Fig. 1 (a) $P(\omega'\geq\omega)$ fitted by SPL function   and   (b) $P(\omega'\geq\omega)$ fitted by exponential function for the 2004 Athens Olympic network.

## 3. University matriculation network

In Chinese universities, colleges or departments are divided into undergraduate specializations. The undergraduates of different specializations are taught by different teaching schemes. In order to enter a specialization, a middle school student has to pass the national matriculation with the total marks higher than a lowest value. The university recruitment process is divided into several (roughly speaking, 6) batches in each province or region. Different batches have different lowest values of the matriculation mark. Different specializations may have the recruitment right in different batches depending on their academic levels. Also, each specialization may have the recruitment right in different batches in different geographical regions. A specialization with a higher batch can recruit higher mark students. Therefore, universities are competing in the recruitment process in many "batches", which depend on specialization and geographical region, to recruit more and better students. From a different view point, all the universities are also cooperating in the total recruitment process to successfully complete the well-organized job. We construct the university matriculation network by defining the batches as the acts, universities as the actors, and the lowest matriculation marks as the node weight, $x_{il}$, respectively. There are 51 batches and 2277 universities. The data were taken from the web stations www.hneeb.cn; www.jszs.net; www.lnzsks.com; and www.nm.zsks.cn.[14,16] Again, we observed that the empirical cumulative distribution of the NNTW, $P(\omega'\geq\omega)$, can be fitted by SPL functions (see Fig. 2a) with the parameters $\alpha=1$ and $\gamma=3684$. The data can be fitted by an exponential function as shown in Fig. 2b.

Typically, the university matriculation lasts for several days during which the universities compete for better students. The act duration data can also be obtained. For example, in 2006, the recruitment processes (from the time when the university delegacies get together to recruit students until the task is accomplished and they go home) of the first batch, the second batch, the third batch and the fourth batch in Shaanxi province lasted for 4 days, 4 days, 3 days and 4 days, respectively. Within all the batches, the longest act duration is 11 days, and the shortest one is 3 days. The average of the act durations for all the batches is around 6 days.

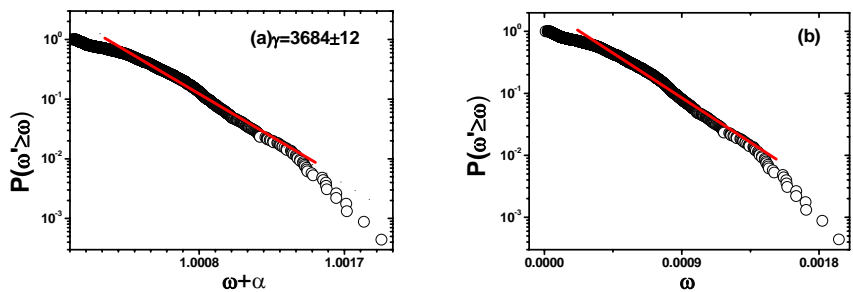

Fig. 2 (a) $P(\omega'\geq\omega)$ fitted by SPL function and (b) $P(\omega'\geq\omega)$ fitted by exponential function for the university matriculation network.

## 4. University independent recruitment network

In China, all the universities can recruit new students from middle schools only via a nationwide uniform activity. All the middle students have to pass the national matriculation. The delegacies of different academic level universities then get together to select middle school students according to the matriculation marks of the students and the university batch rights just as introduced in last section. However, in recent years, some best universities are awarded the right to recruit middle school students before the national matriculation. They can perform their own examinations or interviews and then recruit new students by their own decision. We call this as "independent recruitment". The middle schools can achieve better reputations if more of their students have been recruited by these top level universities. Therefore, the middle schools collaborate to accomplish the independent recruitment activity and simultaneously compete for sending more students to the universities. We define the middle schools as the actors, the independent recruitment universities as the acts, and the number of the recruited students as the node weights. In 2006, there were 53 universities, which had the independent recruitment right. However, only the data of 52 were available. 1546 middle schools and their recruited student numbers were included in the data (from www.chsi.com.cn). [11,14] In Fig.3 we can see that the cumulative distribution of NNTW $P(\omega'\geq\omega)$ can be well fitted by the SPL function with the parameters $\alpha$=0.00058 and $\gamma$=2.696.

It was a pity that we were not being able to get the reliable act duration data of this system.

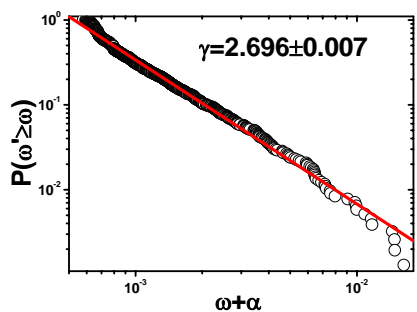

Fig.3: The $P(\omega'\geq\omega)$ for independent recruitment network

## 5. Undergraduate course selection network of Yangzhou University

Yangzhou University (YZU) is a rather new university that was founded in 1992 by the union of 7 smaller colleges. Although its history is short, it develops quite quickly in recent years. Now in its 27 colleges, YZU offers 98 undergraduate programs covering 11 disciplines for 33900 undergraduates. In addition to the courses typically presented in every college, YZU also offers 121 general support courses between 2002 and 2006, which cover all the natural and social scientific disciplines. Every undergraduate has to take at least 4 selective courses among 121 ones. These 121 general support courses can be regarded as the parts of 78 scientific subjects such as physics, mathematic, art and so on. One course may belong to more than one scientific subject. Based on the course selection data of 65,536 undergraduates we built an undergraduate course selection network of YZU.[10,11,14,15] We define the general support courses as the actors. Some certain actors are related to each other by being selected by the students from the same scientific subject (these scientific subjects are defined as the acts of the bipartite network). The node weight is defined as the number of the undergraduates who take this selective general support course. One may say that the general support courses cooperate to form the general education system in YZU, and also compete for attracting more students. Fig.4 shows the cumulative distribution of the NNTW $P(\omega'\geq\omega)$. Again, we can see that the data can be fitted quite well by the SPL function with the fitted parameters $\alpha=0.01$ and $\gamma=3.16$.

The duration data of all the scientific subjects in Yangzhou University can also be obtained. Therefore, up to the data collecting year, i.e., 2006, the longest and shortest act durations are 104 and 8 years, respectively. The average of the act durations is around 55 years.

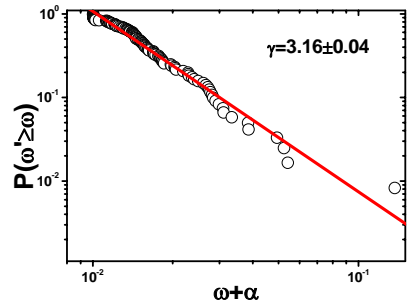

Fig.4: The $P(\omega'\geq\omega)$ of undergraduate course selection network of YZU

## 6. Book borrowing network of YZU library

We also constructed a book borrowing network of YZU library. In library, some books are very popular and have been borrowed frequently by the readers. In comparison, some books have never been borrowed. Such kind of book borrowing information is very important for evaluating the importance of a certain book and is useful for assessing the knowledge configuration of the students. In YZU library, there are 15204 physics books with some books having the same name (for example, for same important subjects, there are usually more than 30 books having the same book name). We collected the borrowing record data for 3207 books with nonzero borrowing record before 2006. These books are further classified into 227 different scientific subjects (such as electromagnetism, quantum mechanics, and so on). One subject includes many different books, and one book may belong to more than one scientific subject. In constructing the bipartite network, we define the books as actors and the scientific subjects as acts. The actors (books) cooperate in a certain act (scientific subject) to provide the necessary knowledge, and also compete for being borrowed more frequently. Therefore, we define the node weight as the number of borrowing records. Totally 88981 records are used in constructing the network. [10,11,14] The cumulative distribution of the NNTW $P(\omega'\geq\omega)$ is presented in Fig.5. The data are well fitted by the SPL function with fitted parameters $\alpha=0.00011$ and $\gamma=2.187$.

We were not being able to get the reliable act duration data of this system.

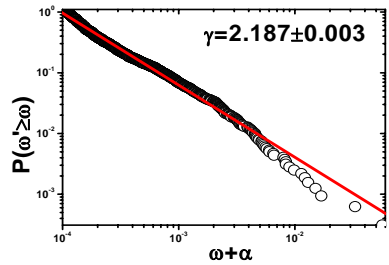

Fig.5: The $P(\omega'\geq\omega)$ for YZU library network

## 7. Training institution network

In recent 20 years Chinese employment market becomes very active. As the result, there have appeared many training institutions which offer variety different training courses. The institutions providing the same training courses cooperate to form the proper market, and also compete for more tuition fee. In the network, the training institutions are defined as the actors and the training courses are defined as the acts. The tuition fee for a certain training course is defined as the node weight. We collected the data of 398 training institutions in China, which are authorized by the national ministry of education and/or the national labor bureau. Until the data collection, 2006, totally 2673 training courses are included in the data (from www.ot51.com, www.00100.cc, www.people.com.cn et al.). [11,14,15] Fig.6 shows the cumulative distribution of the NNTW, $P(\omega'\geq\omega)$, for the training institution network, which can be well fitted by the SPL function with the parameters $\alpha=0.000065$ and $\gamma=2.100$.

It is not easy to get the act duration data for all the acts. However, it is really pointless as discussed in first section. Instead, we randomly collected the act duration data for 10 acts, and then counted the longest, the shortest, and the average values within the sample data. The values are 11, 2 and 7 years, respectively. We believe that the order of magnitudes is reliable, and this is enough for the future dynamics study.

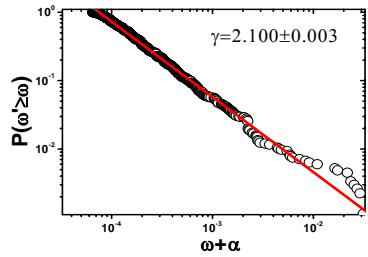

Fig.6 is the $P(\omega'\geq\omega)$ of training institution network

## 8. Supermarket network

Supermarkets selling the same commodities collaborate to provide better services to the customers and compete to attract more buyers for more profits. In a web site, www.dianping.com, the buyers give marks for each commodity in every supermarket according to the supermarket environment, service and each commodity quality and price. For a supermarket, higher marks mean more buyers and profits. We define the supermarkets as the actors, the commodities as the acts, and the buyer's marks as the node weights. Totally 1046 kinds of commodities and 570 supermarkets are included in our data. [14] The cumulative distribution of the NNTW, $P(\omega'\geq\omega)$, is presented in Fig.7 together with the good SPL fitting. The parameters, $\alpha$ and $\gamma$, are 0.011 and 8.21, respectively.

Unfortunately, we were not being able to get the reliable act duration data of this system.

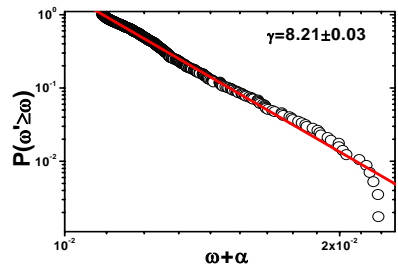

Fig.7: The $P(\omega'\geq\omega)$ for supermarket network

## 9. Information technique product selling network

The market of information technique (IT) products, including mobile phone, computer, digital camera, and so on, is another example which shows typical collaboration-competition characteristics. If different manufacturers produce the same type of IT products, they compete in the selling market. Of course, these manufacturers also collaborate to supply enough IT products, and to maintain the market order. We construct a collaboration-competition bipartite network in which the manufactures are defined as actors and the IT products are defined as acts. On the website www.pcpop.com, there are detailed introductions to each IT product produced by a specific manufacturer. This web site also gives the "attention rank" of the manufacturers for each IT product according to the total browsing time by the customers. To some extent, such "attention rank" is relevant to the competition abilities of the manufactures, and can be used to quantify the competition achievement, i.e., the profit. Therefore, we define the "attention rank" as the node weight of the actor nodes. We collected 265 manufacturers and 2121 IT products from the website. [11,13,14] Fig.8 shows the cumulative distribution of the NNTW, $P(\omega'\geq\omega)$, and its SPL fitting. The parameters are $\alpha=0.00135$ and $\gamma=4.465$.

It is difficult and pointless to determine the act duration for each IT product. We randomly select 10 samples of IT products and find their act duration data. Then we count the shortest, longest and the average values of the act durations within the sample IT products, which are 94, 5 and 28 years, respectively (until 2006, the data collection time). Similarly, we believe the order of magnitudes is reliable.

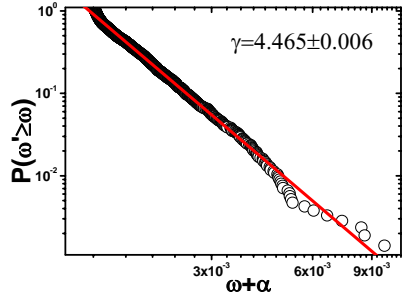

Fig.8: The $P(\omega'\geq\omega)$ for IT product selling network

## 10. Notebook PC selling network at Taobao website

In recent years, on-line shopping by internet becomes more and more popular. Taobao (www.taobao.com) is one of the most famous on-line shopping mall in China. Many shops register in Taobao and sell many kinds of commodities through the Taobao website. Thousands of shops sell notebook PC. The shops collaborate to provide proper notebook PC selling service, and simultaneously compete for more profit. In this bipartite network, the shops are defined as actors, and the notebook PC types are defined as the acts. Usually, the price for the same type of notebook PC can be quite different in different shops. The shops with better reputation can sell out the same type of notebook PCs with higher price. Of course, these shops make more profit. Therefore, the selling price can be defined as the node weight of an actor. Totally 53 notebook PC types and 4711 notebook PC shops from the Taobao on-line shopping mall were collected. [14] Fig. 9 shows the cumulative distribution of the NNTW, $P(\omega'\geq\omega)$, and its SPL fitting. The parameters are $\alpha=0.00088$ and $\gamma=6.063$.

Refs. [17,18] provided the surveys, which let us know that, in average, a type of PC computer is updated every 3 years. Soon after the appearance of a new type of PC computer, the production of the old one stops. Accordingly, the collaboration/competition between the on-line shops for selling the old product terminates. Therefore, we believe that the reliable average act duration of a notebook PC product is 3 years. This should be correct in order of magnitudes.

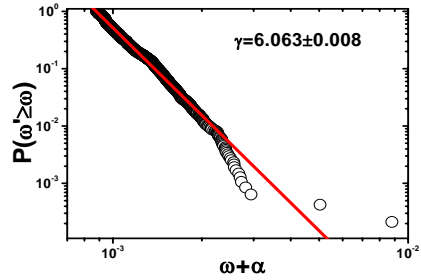

Fig. 9: The $P(\omega'\geq\omega)$ for taobao notebook PC selling network

## 11. Beijing restaurant network

Chinese food is famous in the world. Usually, one Chinese restaurant serves a large number of different cooked dishes, and the same cooked dish can be served by many restaurants. We constructed a collaborate-competition bipartite network of the restaurants in Beijing, in which the restaurants are defined as actors and the cooked dishes are defined as acts. In addition to collaborating to provide the food services, the restaurants serving the same type of dishes also compete to attract more customers and therefore earn more profits. The competition ability of each restaurant can be represented by the "attention degree" which is quantified by the customer marks given in the Dianping website (www.dianping.com/beijing) for each dish. We define the customer marks as the node weight. Until 2006, we collected 688 cooked dishes in 3337 restaurants. [11,14,15] Fig.10 shows the cumulative distribution of the NNTW, $P(\omega'\geq\omega)$, of the Beijing restaurant network. The parameters are $\alpha$=0.0000365 and $\gamma$=2.118.

Again, we randomly select 10 samples and find their act duration data. Then we count the shortest, longest and the average values of the act durations within the samples, which are 159, 9, 75 years until 2006, respectively. It should be reliable in the order of magnitudes.

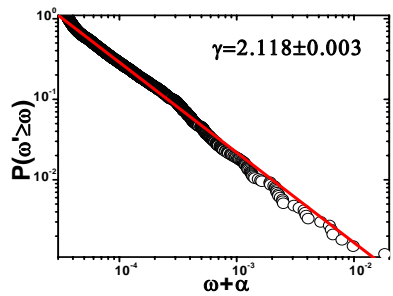

Fig.10: The $P(\omega'\geq\omega)$ of Beijing restaurants network

## 12. Mixed drink network

The mixed drinks (such as cocktails) usually contain a large number of ingredients according to the consumer's taste, and many mixed drinks may share the same ingredients. We construct the mixed drink network by defining the component ingredients as actors and the mixed drinks as acts. The ingredients collaborate to form mixed drinks with different tastes. Simultaneously, the ingredients contained in a common mixed drink can be regarded as being competing since the ingredients have different relative importance. As the first step investigation, we very simply suppose that a certain ingredient in higher proportion is relatively more important. Therefore, we define the relative proportion of each ingredient in a certain mixed drink as the node weight. Until 2006, we collected 7804 mixed drinks and 1501 ingredients. The proportions of the ingredients in each mixed drink are also obtained (www.drinknation.com). [11,12,14] Fig.11 shows the cumulative distribution of the NNTW, $P(\omega'\geq\omega)$, of the mixed drink network and the SPL fitting. The parameters are $\alpha=0.000051$ and $\gamma=1.783$.

Similarly, we randomly select 7 samples and find their act duration data. Then we count the shortest, longest and the average values of the act durations within the samples, which are 500, 100, and 340 years, respectively. It should be reliable in the order of magnitudes.

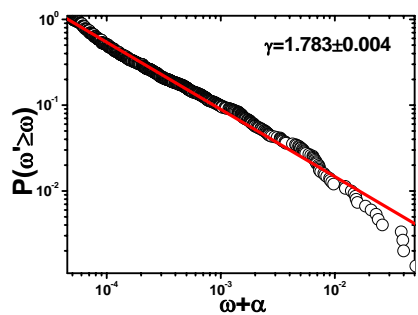

Fig.11 : The $P(\omega'\geq\omega)$ of mixed drink network

## 13. China mainland movie network

The data of China mainland movies are taken from the database (http://www.movdb.com). Based on the data, we constructed a China mainland movie network in which the movies are defined as actors and the leading movie actors as acts. Such definition is different from that in Refs. [2,20,21]. The movies acted by the same leading actors often belong to a common movie type, which has a certain type of audience. In this sense, these movies compete to attract more audience in addition to collaborating to provide abundant entertainments. More downloading numbers of a movie indicates more audience and therefore more ticket office income. We therefore define the downloading numbers as the node weight. We collected the data for 3084 movies involving 920 leading actors from the database before April 29, 2007. [11,14,15,19] Fig.12 shows the cumulative distribution of the NNTW, $P(\omega'\geq\omega)$, of China mainland movie network. The SPL fitting parameters are $\alpha=0.0000161$ and $\gamma=1.662$.

Again, we randomly select 10 samples and find their act duration data. Then we count the shortest, longest and the average values of the act durations within the samples, which are 45, 13 and 26 years, respectively (until 2006, the data collection). It should be reliable in the order of magnitudes.

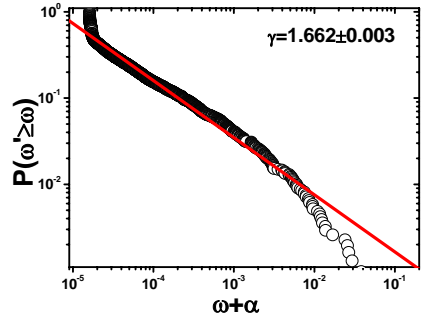

Fig.12: The *P(ω'≥ω)* of China mainland movie network

## 14. World language distribution network

Based on the data from Ethnologue website (http://www.ethnologue.com, the 15th edition, published in 2005) which shows the populations speaking each language in a country (or a relatively independent region), we constructed a world language distribution network. The languages are defined as actors, and the countries are defined as acts. In the bipartite graph, if a certain language is used in a country, this language and the country are connected by an edge. In the projected unipartite graph, the edge between two languages means that the languages are used simultaneously at least in one common country. In a very long time consideration, the languages can be considered as being competing to be used by more people. As the result, some languages have died out, but some other languages have been spoken by more and more people and spread to more and more geographical regions. The languages also collaborate in the common regions to accomplish the communications between the people. Therefore, we define the populations speaking a certain language in the country as the node weight. We collected 6142 kinds of languages used in 236 countries and regions.[22] The total number of the language speakers is $5.2385 \times 10^9$. Since the Ethnologue data are taken from multiple sources, the sum of the languages' population may not exactly equal to the total population in the world. [11,14] Fig.13 shows the cumulative distribution of the NNTW, *P(ω'≥ω)*, for the word language network and its SPL fitting. The parameters are *α*=0.0000014 and *γ*=1.6316.

Since this is the language network, an act is defined as a country/region with its legal languages or the languages spoken by most of the people unchanged. If the languages basically change in the oral or written forms, we define that the old act dies out and a new act comes into the world. Chinese language ("Han" language, which has been used by most of Chinese people) has no basic change in about 5000 years although its oral and written forms have been developing in such a long period of time. As we know, this is the longest act duration in this network. One can mention quite some samples in which political or other factors induce the oral or written form basic change of the legal languages or the languages spoken by most of the people in a quite short time. For example, Singapore was governed by England for a long time and also by Japan and Malaysia for short time periods. However, the legal language was always English until 1965, when Singapore declared its autocephaly. The legal language changed to English, Chinese (Mandarin), Malay and Tamil since then. So, the act duration of Singapore is 40 years (until 2005). As an even better example, in Vietnam, most of the people used Chinese as the written language for a long time, the presently used alphabetic writing language was created only 65 years ago during the France colonial domination. [23] We believe that both the act duration data (they are same in the order of magnitudes) represent the shortest act duration of the world language network.

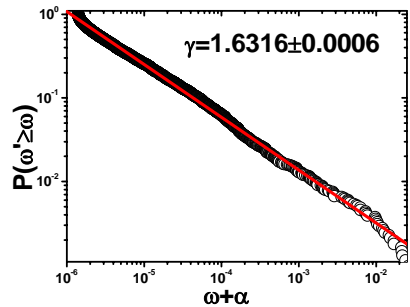

Fig.13: The *P(ω'≥ω)* for the word language distribution network

## 15. Human acupuncture point network

Acupuncture is one of the most important traditional Chinese therapy means. Because of its practical efficacy in curing some diseases, acupuncture becomes more and more popular over the world. There are 187 key acupuncture points in human body. To treat a certain disease, several specific acupuncture points need to be acupunctured simultaneously. In this sense, the acupuncture points collaborate in curing a certain disease. Therefore, we can define the diseases treated by acupuncture as acts, and the acupuncture points as actors. However, for a certain disease, not all of the relevant acupuncture points play the equal role. Some acupuncture points are more crucial. Usually, these relatively more important acupuncture points, in treating a certain disease, need more acupuncture times. Therefore, we define the acupuncture time as the node weight. Totally 108 different kinds of diseases and 187 acupuncture points are involved in our data (from www.acutimes.com ). [11,14] Again, as shown in Fig.14, the cumulative distribution of the NNTW, *P(ω'≥ω)*, shows a SPL distribution with the parameters *α*=0.0045 and *γ*=2.552.

We were not being able to get the reliable act duration data of this system.

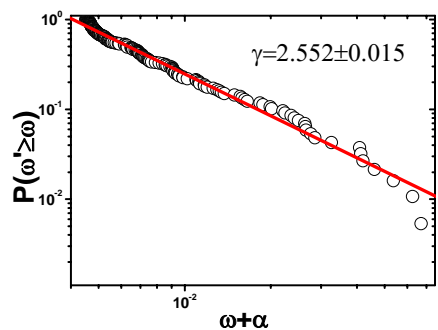

Fig.14: The *P(ω'≥ω)* for Human acupuncture point network

## Acknowledgement

The studies were supported by the National Natural Science Foundation of China under Grant Nos 10635040 and 70671089.

## Appendix 1

SPL distribution of the NNTW can be expressed as $P(\omega) \propto (\omega+\alpha)^{-\gamma}$, where $0<\omega_i<1$ and $\sum_{i=1}^{N} \omega_i = 1$. $N$ is the total number of the nodes. In general, $N$ is thousands upon thousands. So we usually have $0<\omega_i<<1$. When $\alpha$ approaches to 1, we have $\alpha>>\omega_i$, namely $\omega_i/\alpha \to 0$. Changing the SPL function, we obtain that

$$P(\omega) \propto (\omega+\alpha)^{-\gamma} = \alpha^{-\gamma}(1+\frac{\omega}{\alpha})^{-\gamma} = \alpha^{-\gamma}[\ (1+\frac{\omega}{\alpha})^{\frac{\alpha}{\omega}}]^{(-\gamma)\cdot\frac{\omega}{\alpha}}$$

When α approaches to 1, by using the formula $\lim_{x \to 0}(1+x)^{\frac{1}{x}} = e$, we obtain

$$P(\omega) \propto \alpha^{-\gamma} e^{(-\gamma)\cdot\frac{\omega}{\alpha}} = \alpha^{-\gamma} e^{-\frac{\gamma}{\alpha}\cdot\omega} \propto (e^{\frac{\gamma}{\alpha}})^{-\omega} \propto c^{-\omega}$$

So we know that SPL function tends to the exponential function when α approaches to 1.